\documentclass[%
 reprint,
 amsmath,amssymb,
 aps,
]{revtex4-2}

\usepackage{graphicx}
\usepackage{dcolumn}
\usepackage{bm}

\usepackage[colorlinks,citecolor=green,linkcolor=red,urlcolor=blue,bookmarks=false,hypertexnames=true]{hyperref}

\begin{document}

\preprint{APS/123-QED}

\title{Squigglebot: a battery powered spherical rolling robot\\ as a model active matter system to measure its energetics}

\author{Soumen Das}
\email{soumend80@gmail.com}
\author{Anit Sane}
\author{Shankar Ghosh}%
\affiliation{ Department of Condensed Matter Physics and Materials Science,\\ Tata Institute of Fundamental Research, Mumbai 400005, India}





\date{\today}

\begin{abstract}
Active matter systems use their internal or ambient source of energy and dissipate them at the scale of individual constituent particles to generate motion. Direct measurement of the energy influx for individual particles has not been achieved in the experiments. Here we present ``Squigglebot" - a battery powered spherical rolling robot based on open source hardware as an artificial active matter system whose energy consumption as well as the energy dissipation into different modes of motion both can be measured experimentally. This can serve as a prototype system to study a number of interesting problems in non-equilibrium statistical physics, where details of the energetics are required.
\end{abstract}

\maketitle

\section{\label{sec:Introduction}Introduction}
The study of active matter systems whose constituent particles consume and dissipate energy in order to generate mechanical forces and move in a collective manner has been an area of  interest in the field of non-equilibrium statistical physics \cite{ramaswamy2017active,ramaswamy2010mechanics_annurev-conmatphys}. These are mostly observed in the biological settings which include all living organisms; spatial organization of chromatin in the nucleus \cite{agrawal2017chromatin}, self organization of microtubules \cite{ndlec1997self} and actin present in the cyctoskeleton of living cells \cite{juelicher2007active}; animal groups \cite{ballerini2008interaction}  (e.g. bird flocks, fish schools, herd of sheep's) etc. Artificial active matter include self propelled particles \cite{vicsek1995novel}, driven granular systems \cite{deseigne2010collective}, swarming robots \cite{wang2021emergent} etc. Sustained consumption and dissipation of energy make these systems inherently out of equilibrium. In most studies of active matter systems the energy input at the level of individual particles are not monitored directly. Here we introduce ``Squigglebot" - a battery powered rotating spherical robot as a model active matter particle whose energy intake as well as the energy expenditure both can be measured experimentally.  

This can be useful in situations where details of both input and output energies are needed in non-equilibrium systems. For example one can study the energy efficiency of an individual active particle, i.e., how it converts its internal energy to achieve its desired motion. Similarly, how active matter systems move away from equilibrium is an interesting question which has not been studied beyond a few theoretical models \cite{Fodor2016PRL}. These approaches use the idea of entropy production based on the statistics of individual microscopic trajectories which is hard to obtain in experiments. However one can construct an equivalent measure of entropy production based on the distributions of the input energy in the system. 

\section{\label{sec:Design}Design}

\subsection{Mechanical Construction}

\begin{figure}
\includegraphics[width=.9\linewidth]{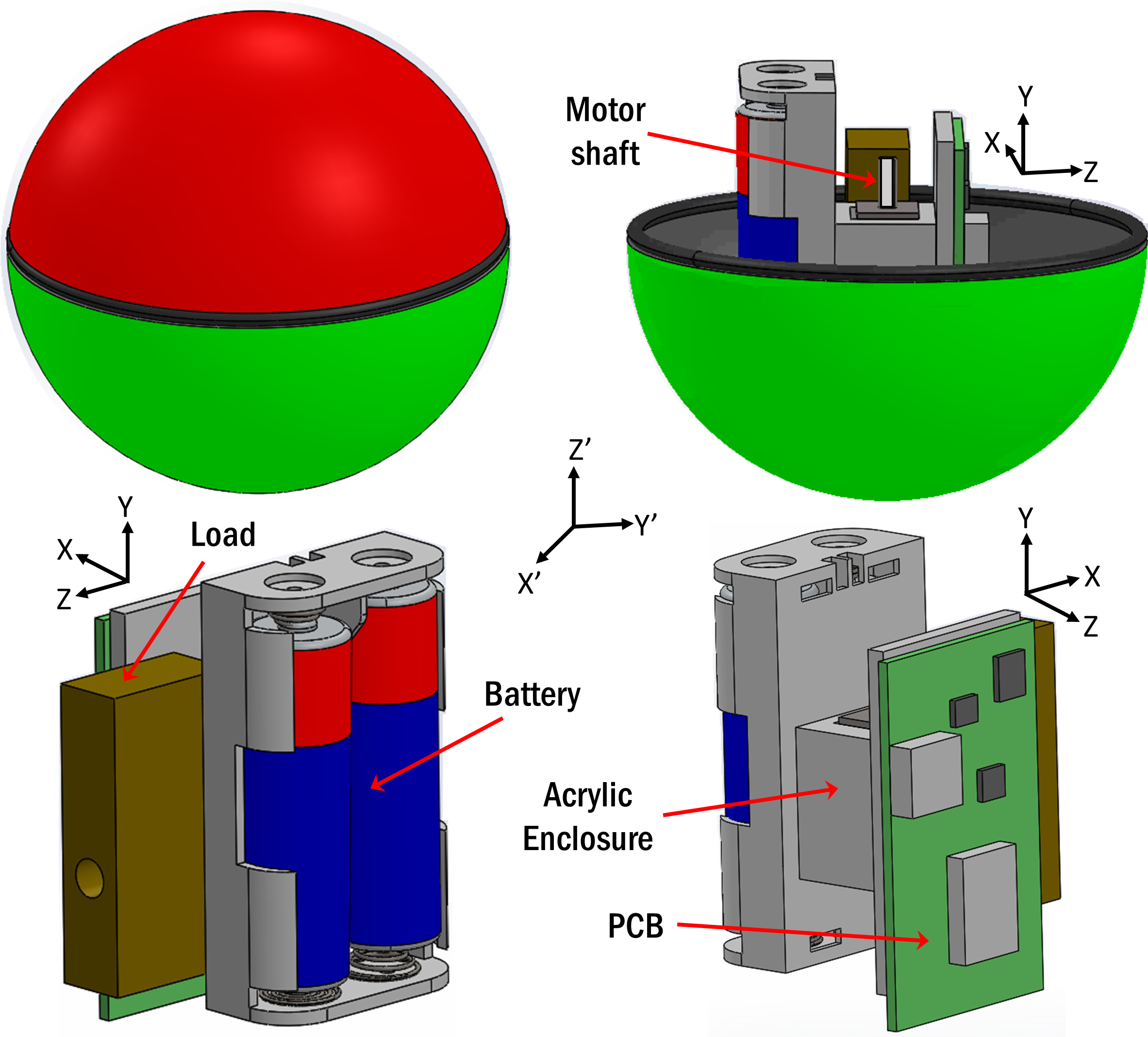}
\caption{Schematic of the mechanical assembly with its different components pointed out. X-Y-Z is the body frame and X'-Y'-Z' is the lab frame. The motor rotates about the body Y axis.}
\label{fig:mech_assembly}
\end{figure}

\begin{figure*}[t]
\includegraphics[width=.75\linewidth]{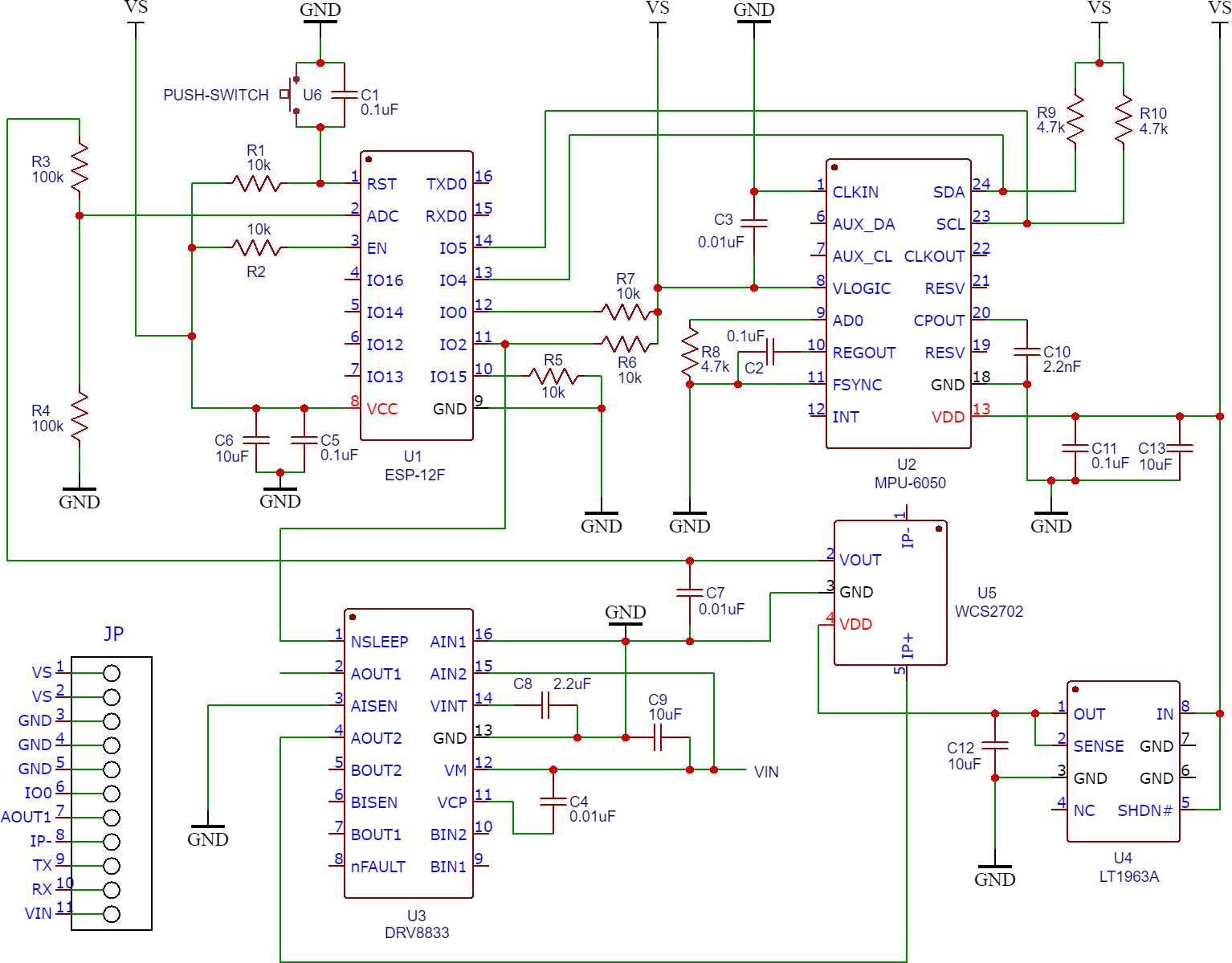}
\caption{\label{fig:schematic} Schematics of the circuit diagram used in the Squigglebot. JP indicate the pins for connecting jumper cables to the pcb.}
\end{figure*}

The experimental system, based on Squiggle-ball toy, consists of a plastic spherical shell of diameter $80~\rm{mm}$ inside which a dc motor is mounted along the axis. Shaft of the motor is connected to the shell. The body of the motor is held in an acrylic enclosure. On one side of the motor, a solid block of brass of dimension $10\times20\times40~\rm{mm}$ and mass $57~\rm{gm}$ is screwed to the acrylic. A $2\times\rm{AA}$ battery holder case and a printed circuit board (pcb) is attached to the adjacent two sides of the acrylic enclosure. Due to the asymmetric mass distribution of the brass load - motor assembly, the center of mass of the ball changes as the motor rotates. When placed on a solid surface with the motor turned on, the sphere rolls in a chaotic manner \cite{ilin2017dynamics}. When the sphere collides with a wall or any other rigid object, it rebounds in a random direction. The schematic of the mechanical assembly of the ``Squigglebot'' made using SolidWorks is shown in Fig.~\ref{fig:mech_assembly}. 
The moment of inertia of the assembly in the body frame can be calculated using SolidWorks as  

$J = \begin{pmatrix}
730.58 & 12.02 & -12.27\\
12.02 & 717.19 & 22.27\\
-12.27 & 22.27 & 698.53 \end{pmatrix}$ gm * $\rm{cm}^2$.


\subsection{Electronics}


The schematic diagram of the electronics circuit is shown in Fig.~\ref{fig:schematic}. The micro-controller used for control, data collection and sensing is ESP-12F from Ai-Thinker Technology \cite{esp-12f}. It is an Wi-Fi enabled microchip based on ESP8266 SoC (System on Chip) module. ESP8266 integrates Tensilica’s L106 Diamond series 32-bit RISC processor running at 80 MHz, 36 KB of RAM, 4 MB of external SPI flash, 10 bit ADC (analog to digital converter), I/O and a PCB-on-board antenna in the same chip. 
The module supports standard IEEE 802.11 b/g/n 2.4 GHz Wi-Fi and complete TCP/IP protocol stack. It also supports SDIO, SPI, $\rm{I}^2 \rm{C}$, $\rm{I}^2 \rm{S}$ and UART protocols for serial communication. 

The recommended power supply voltage for its operation is 3 - 3.7 V. Two decoupling capacitors of capacitance 0.1 $\mu \rm{F}$ (C5) and 10 $\mu \rm{F}$ (C6) are placed between the power supply (Vcc) pins and the ground close to the chip to filter out the high and low-frequency noise respectively. A push button switch is connected between the RST pin and ground for resetting the chip manually. RST, EN, GPIO0 and GPIO2 pins are connected to power supply VS via $10~\rm{k}\Omega$ pull-up resistors (R1, R2, R7, R6). GPIO15 is grounded via a $10~\rm{k}\Omega$ pull-down resistor (R5). GPIO0 pin needs to be shorted to ground temporarily to put it into the flash mode for uploading the code via UART protocol using an external programmer chip. 


To measure the instantaneous angular velocity as well as the linear acceleration of the ball, MPU-6050 which is a MEMS based IMU (inertial measurement unit), is used \cite{mpu6050,mpu6050registermap}. MPU-6050 is an integrated 6-axis motion-tracking device that combines a 3-axis gyroscope, 3-axis accelerometer, and a Digital Motion Processor (DMP). A MEMS accelerometer consists of a proof mass suspended on a spring.
When the acceleration is applied on a particular axis, it causes the proof mass to shift to one side. Due to this deflection the capacitance between fixed plate and plate attached to the proof mass is changed. This change in capacitance is proportional to the acceleration. The sensor processes this change in capacitance and converts it into an analog output voltage \cite{nihtianov2018smart}. A MEMS gyroscope sensor is composed of a proof mass which is kept in a continuously oscillating motion. When a rotation is applied, the Coriolis force acting on the moving proof mass changes the direction of the vibration. This causes a capacitance change proportional to the angular velocity which is picked up by the sensing element and then converted to a voltage signal \cite{nihtianov2018smart}.
 
 MPU6050 can measure the angular velocities with four programmable full scale ranges of $\pm 250^{\circ}\rm{/s}$, $\pm 500^{\circ}\rm{/s}$, $\pm 1000^{\circ}\rm{/s}$ and $\pm 2000^{\circ}\rm{/s}$. Similarly it can measure the accelerations with full scale ranges of $\pm 2\rm{g}$, $\pm 4\rm{g}$, $\pm 8\rm{g}$ and $\pm 16\rm{g}$, where $\rm{g}$ is the acceleration due to gravity. We use full scale ranges of $\pm 2000^{\circ}\rm{/s}$ and $\pm 2\rm{g}$ for gyroscope and accelerometer respectively. Following capacitors are used - (i) 0.1 $\mu \rm{F}$ regulator filter capacitor (C2) between REGOUT and ground, (ii) 0.01 $\mu \rm{F}$ bypass capacitor (C3) between VLOGIC and ground, (iii) 2.2 $\rm{nF}$ charge pump capacitor (C10) between CPOUT and ground and (iv) 0.1 $\mu \rm{F}$ bypass capacitor (C11) between power supply VDD and ground pins. $\rm{I}^2 \rm{C}$ \cite{frenzel2015handbook} is used for communication between MPU6050 and ESP-12F micro-controller. SCL is the $\rm{I}^2 \rm{C}$ clock pin used to carry the timing signal supplied by the bus master device (ESP-12F). This pin is connected to the GPIO5 pin on ESP-12F. SDA is the $\rm{I}^2 \rm{C}$ data pin used for both transmitting and receiving data. It is connected to the GPIO4 pin on ESP-12F. Both SCL and SDA pins are connected to the power supply voltage VS via two $4.7~\rm{k}\Omega$ pull-up resistors. The AD0 pin determines the $\rm{I}^2 \rm{C}$ address of the module. This pin is connected to ground using a $4.7~\rm{k}\Omega$ pull-down resistor, which sets its $\rm{I}^2 \rm{C}$ address as $\text{0x68}$ in hexadecimal representation. Data is collected from MPU6050 at 25 Hz which is then transmitted over Wi-Fi by the micro-controller.    


The maximum current that each GPIO pin of ESP8266 can supply is about 6 mA which is not sufficient to drive the motor. Hence, DRV8833 dual H-bridge motor driver is used to drive the motor \cite{drv8833}. Another reason for using the motor driver is that without it, the noise from the motor can reset the control circuitry or burn out internal components of the micro-controller. The capacitors used are - (i) 10 $\mu \rm{F}$ ceramic bypass capacitor (C9) between the device power supply (VM) and ground pins, (ii) 0.01 $\mu \rm{F}$ X7R seramic capacitor (C4) between high-side gate driver VCP and VM pins and (iii) 2.2 $\mu \rm{F}$ bypass capacitor (C8) between VINT and ground pins. AIN1 and AIN2 pins are connected to ground (Logic Low) and power supply VIN (Logic High) respectively. GPIO2 pin of ESP-12F is connected to NSLEEP (enable) pin of the motor driver. Therefore by setting the GPIO2 pin to logic High and Low, the motor driver and in turn the motor can be turned on and off.

\begin{figure}
\includegraphics[width=.9\linewidth]{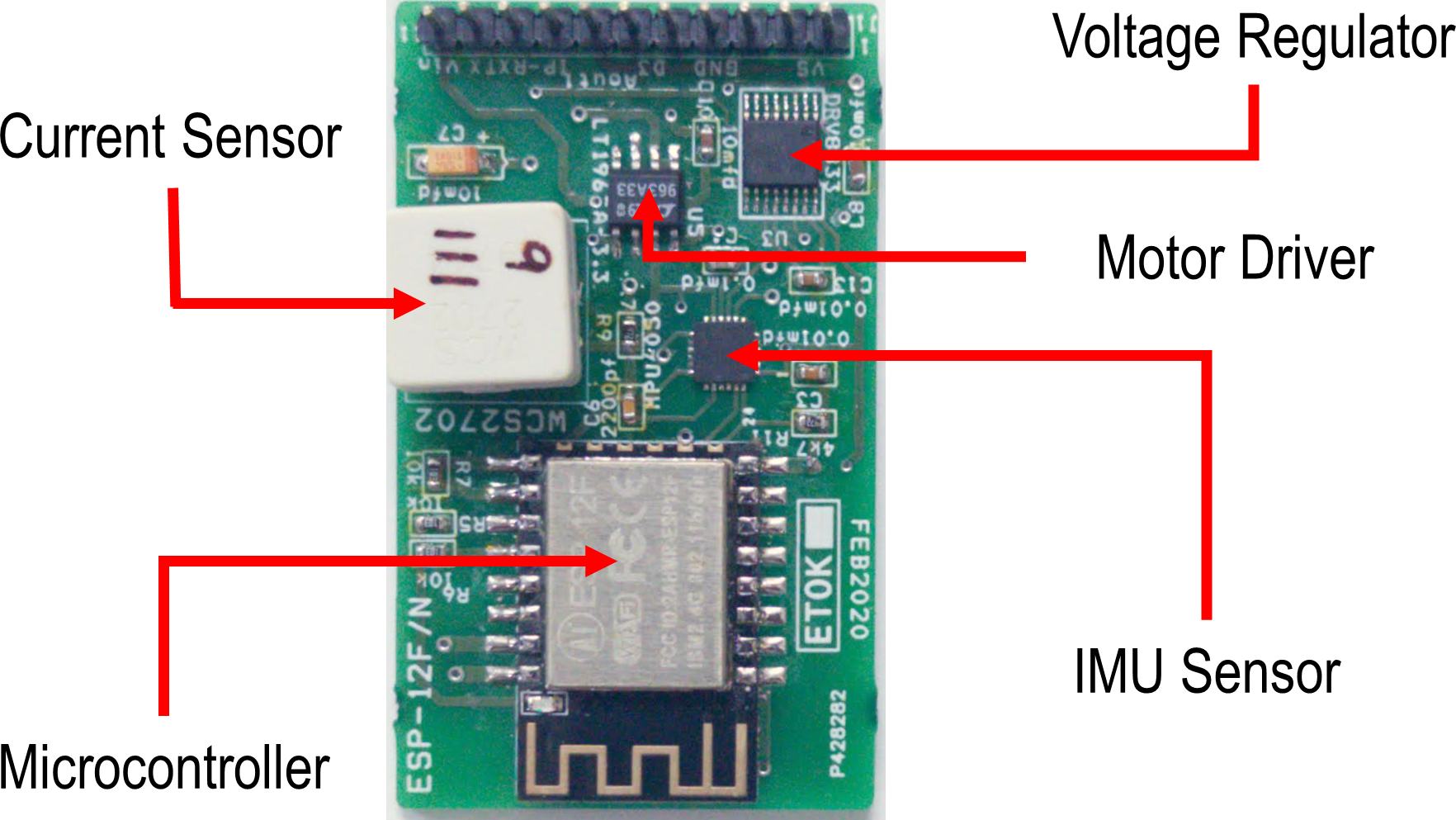}
\caption{\label{fig:pcb} Printed circuit board with the main components pointed out.}
\end{figure}

In order to measure the current drawn by the motor from the battery, a Hall effect based current sensor WCS2702 is used \cite{wcs2702}. A 0.01 $\mu \rm{F}$ capacitor (C7) is connected between the output pin (VOUT) and the ground to reduce the noise. The motor is connected between the IP- pin and AOUT1 pin of the motor driver. IP+ pin is connected to the AOUT2 pin of the motor driver. Here the current flows from IP+ to IP- pin. 

The input power supply of the current sensor (VDD) is kept fixed at 3.3 V using LT1963A-3.3 linear and low-dropout (LDO) voltage regulator \cite{lt1963a}. A 10 $\mu \rm{F}$ bypass capacitor (C13) is connected between the input (IN) and ground pins. Another 10 $\mu \rm{F}$ capacitor (C12) is connected between output (OUT) and ground to prevent oscillations of the output voltage and make it stable.

ADC pin of ESP-12F can handle a maximum voltage of 1 V. Since the output of the current sensor (VOUT) is more than 1 V, a voltage divider (consisting of resistors R3 and R4 - both $100~\rm{k}\Omega$) is used to decrease the voltage of VOUT pin before feeding it to ADC pin. This sets the current resolution of the present experimental setup at 2.3 mA. By reading the voltage of the ADC pin, the current drawn by the motor can be estimated.

Any 3-6 V geared DC motor with a cross section of $10\times12~\rm{mm}$ and D-shaped gearbox output shaft of diameter $3~\rm{mm}$ can be used. For the data presented in this paper, a 60 rpm motor is used with a output shaft of length $9~\rm{mm}$. The no load and stall current are about $10~\rm{mA}$ and $1~\rm{A}$ respectively. The rated and stall torque are $200$ gm-cm and $1600$ gm-cm respectively.


Two 3.7 V 800 mAh Li-ion batteries are used as power supplies. One (VIN) is used to power the motor along with the motor driver and the other one (VS) is used for powering the rest of the electronics including the micro-controller. Ground pins of both batteries as well as all IC's are connected together. The designed printed circuit board is shown in Fig.~\ref{fig:pcb}.


\begin{figure}
\includegraphics[width=.8\linewidth]{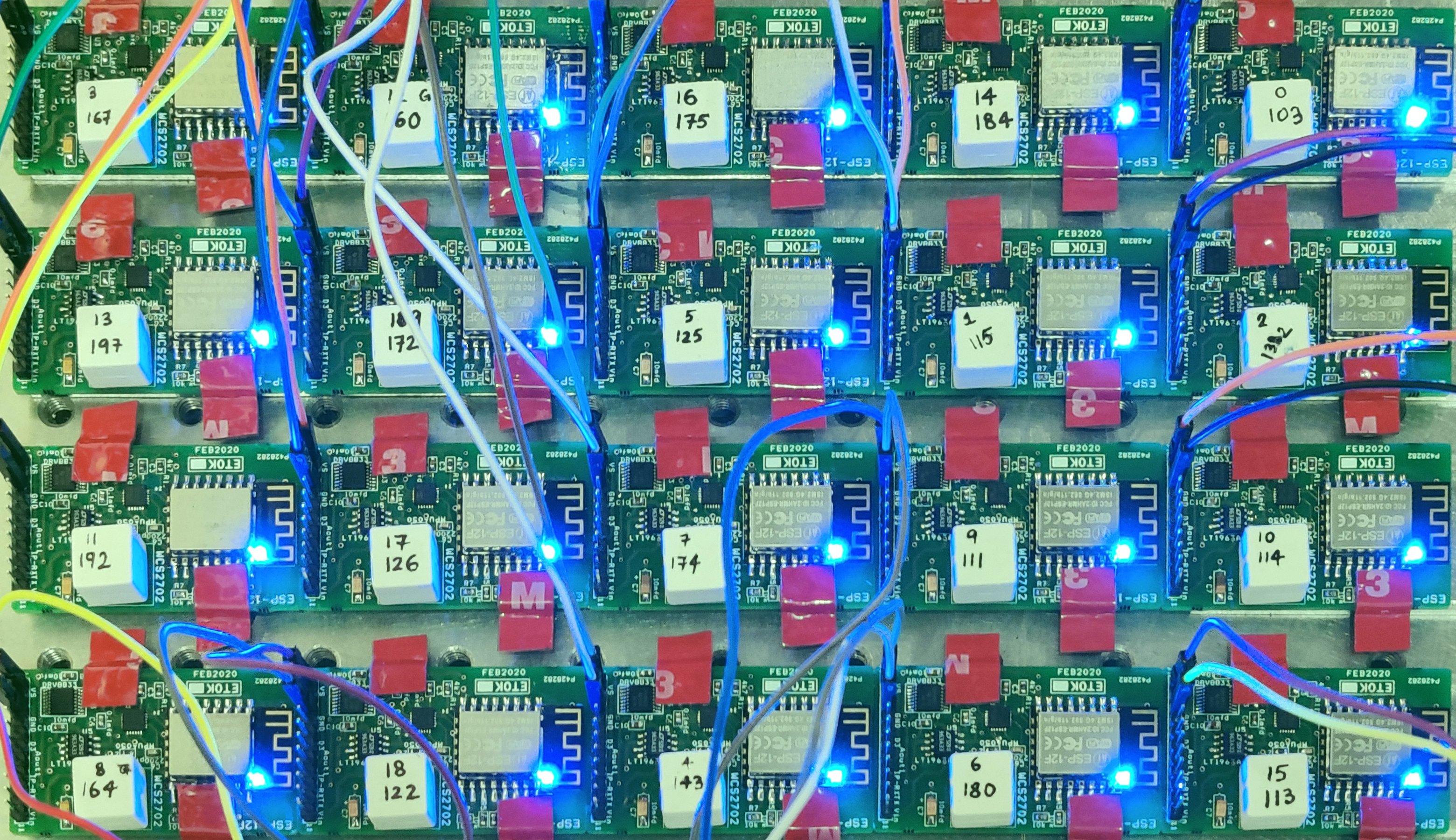}
\caption{\label{fig:calibration} Calibration of accelerometers and gyroscopes in MPU6050s is done by placing them on a horizontal plane.}
\end{figure}

\section{\label{sec:Measurements}Measurements}

We are interested in the energetics of the ``Squigglebot''. Here input energy is the electrical energy consumed by the motor from the battery. On the other hand, output energy consists of the translational and rotational kinetic energies of the ball as well the energy dissipated in the motor and that due to the friction between the ball and the surface on which it is moving. During a single run of the experiment ($\sim$ 2 h 15 min), voltage $V$ of the battery powering the motor almost remains constant ($\sim$ 3.6-3.7 V). Hence monitoring the current drawn by the motor, ${I(t)}$, is enough to estimate the electrical energy which is ${E_e(t)}=\int V*I(t)~dt$. 

 Accelerometer measures the dynamic linear acceleration resulting from motion of a body, as well as the static acceleration due to gravity acting on it. Here the ball has a dynamic acceleration much smaller than the acceleration due to gravity. Since the accelerometer attached to the ball itself is rotating, it becomes difficult to filter out the gravity component and get only the linear acceleration due to the motion of the ball. Hence to keep things simple, we take images of the ball from the top of the arena where it is moving and by detecting the position, we can calculate its linear velocity $\vec{v}$ and the translational energy, $E_T=\frac{1}{2}m|\vec{v}|^2$, where $m$ is the mass of the ``Squigglebot" ($\sim 145~\rm{gm}$). 

Gyroscope measures the angular velocity of the ball with respect to the body frame. By fusing the acceleration and angular velocity from the IMU sensor, one can estimate the rotation matrix $R$ describing the orientation of the body frame with respect to the lab frame \cite{foxlin1996inertial,madgwick2011orientation}. 
It can be used to convert physical quantities from body frame to lab frame. Since energy is invariant in different frames, estimating the rotational kinetic energy in the body frame is sufficient which can be calculated as $E_R=\frac{1}{2}\bm{\vec{\omega}}^T \bm{\tilde{J}} \bm{\vec{\omega}}$, where $\bm{\vec{\omega}}$ is the angular velocity column vector with the components $\omega_x,~\omega_y,~\omega_z$. 

\begin{figure}[t]
\includegraphics[width=.8\linewidth]{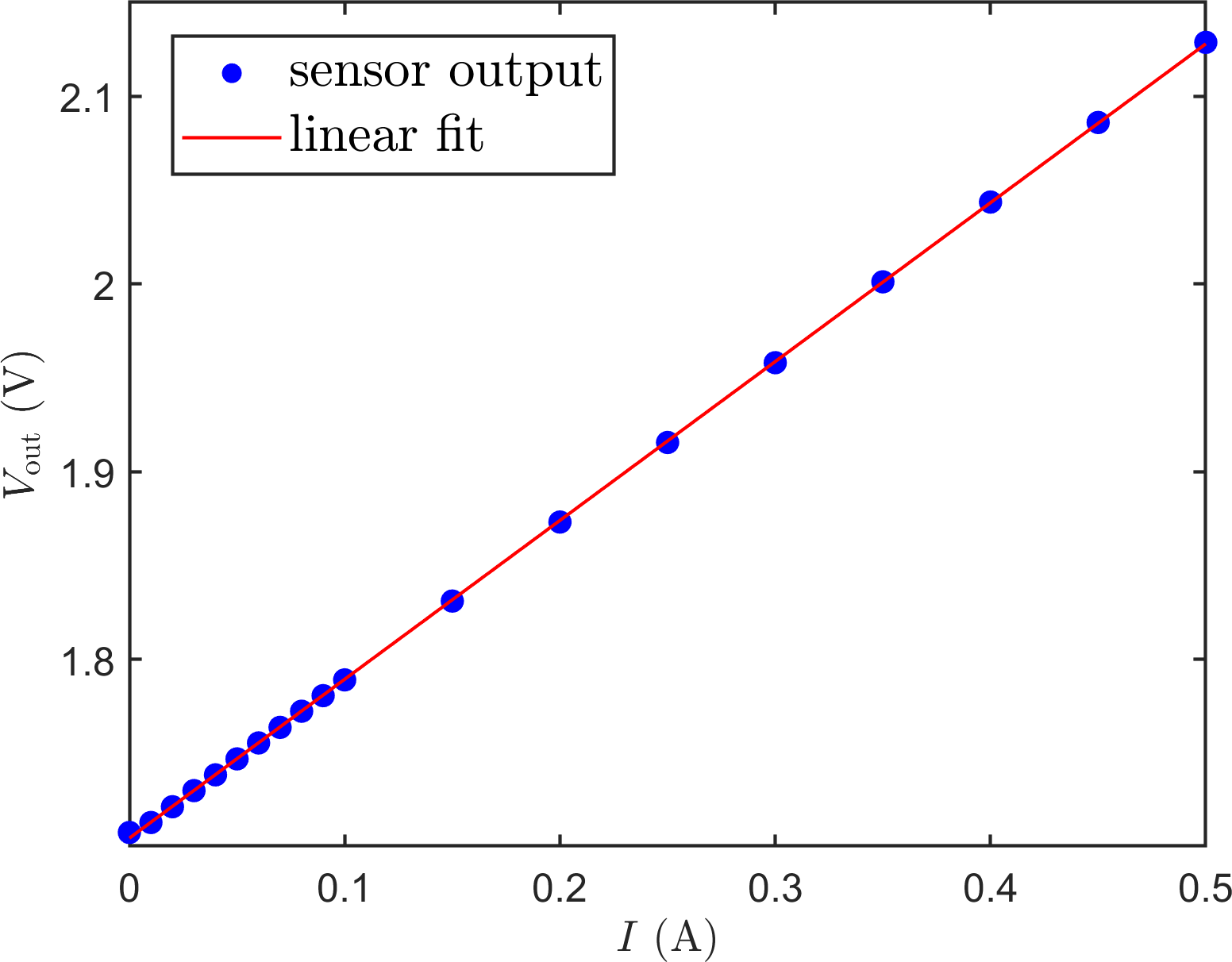}
\caption{\label{fig:current sensor calibration} Calibration curve of the current sensor WCS2702.}
\end{figure}

\begin{figure}[b]
\includegraphics[width=.6\linewidth]{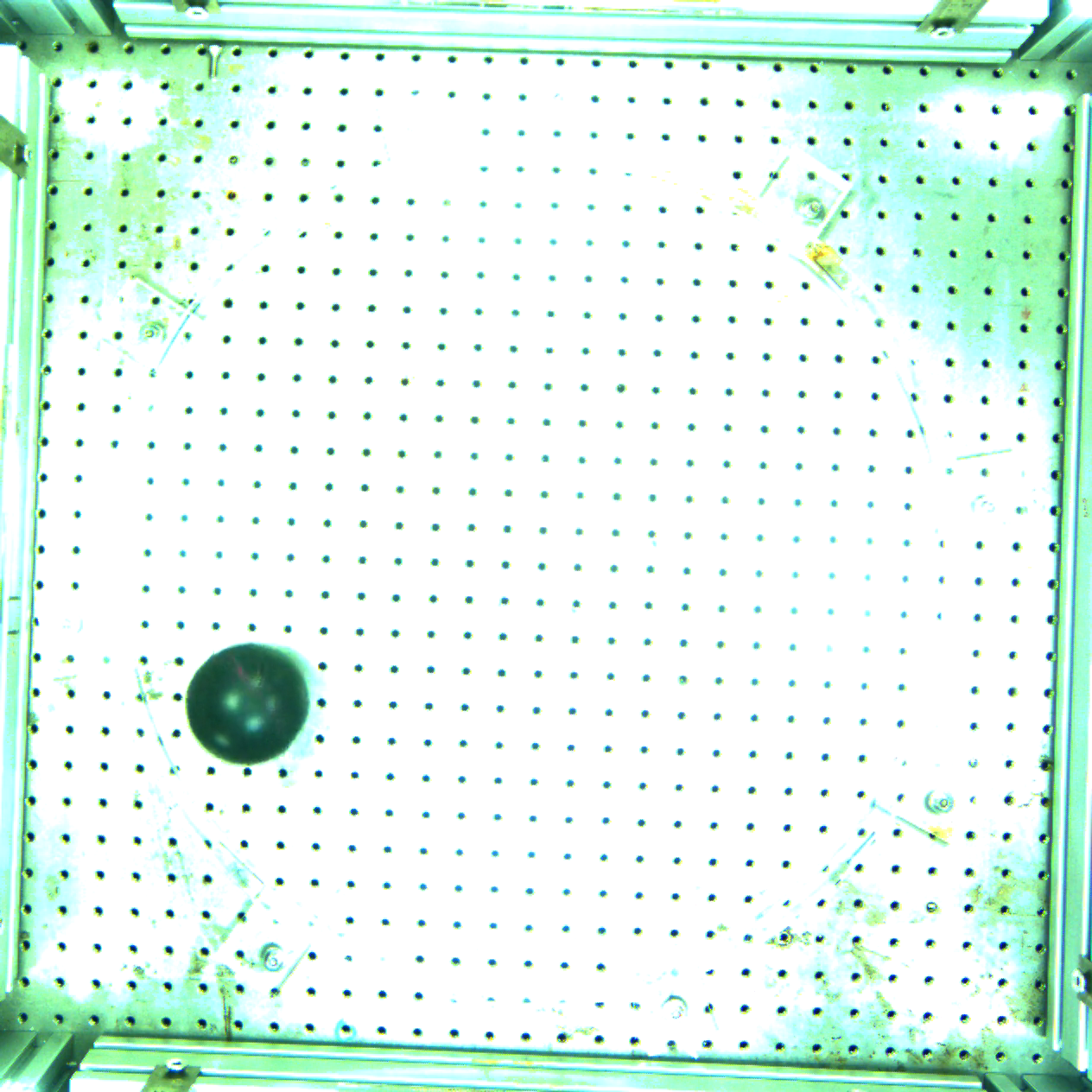}
\caption{\label{fig:img} Snapshot of the ball moving on a horizontal surface.}
\end{figure}

\subsection{Gyroscope and Accelerometer Calibration}
For calibrating the gyroscope and the accelerometer values, the pcbs are sticked using double-sided tape to a flat acrylic plate placed on a horizontal plane and their values are measured (Fig.~\ref{fig:calibration}). These values are then stored in the code as offsets and during the experiment these offsets are subtracted from the measured raw values before transmitting the data. 


\subsection{Current sensor Calibration}
The current sensor generates a voltage proportional to the current flowing through it. It is calibrated using Keithley 2410 sourcemeter as a constant current source. The calibration curve is linear with $V_{\rm{out}}=1.7047+0.8464*I$ as shown in Fig.~\ref{fig:current sensor calibration}. Here $V_{\rm{out}}$ is the output voltage of the current sensor and $I$ is the current flowing through it from IP+ to IP- pin (see Fig.~\ref{fig:schematic} for the direction of current flowing). Thus it has a sensitivity of 0.8464 V/A.

\begin{figure}
\includegraphics[width=.8\linewidth]{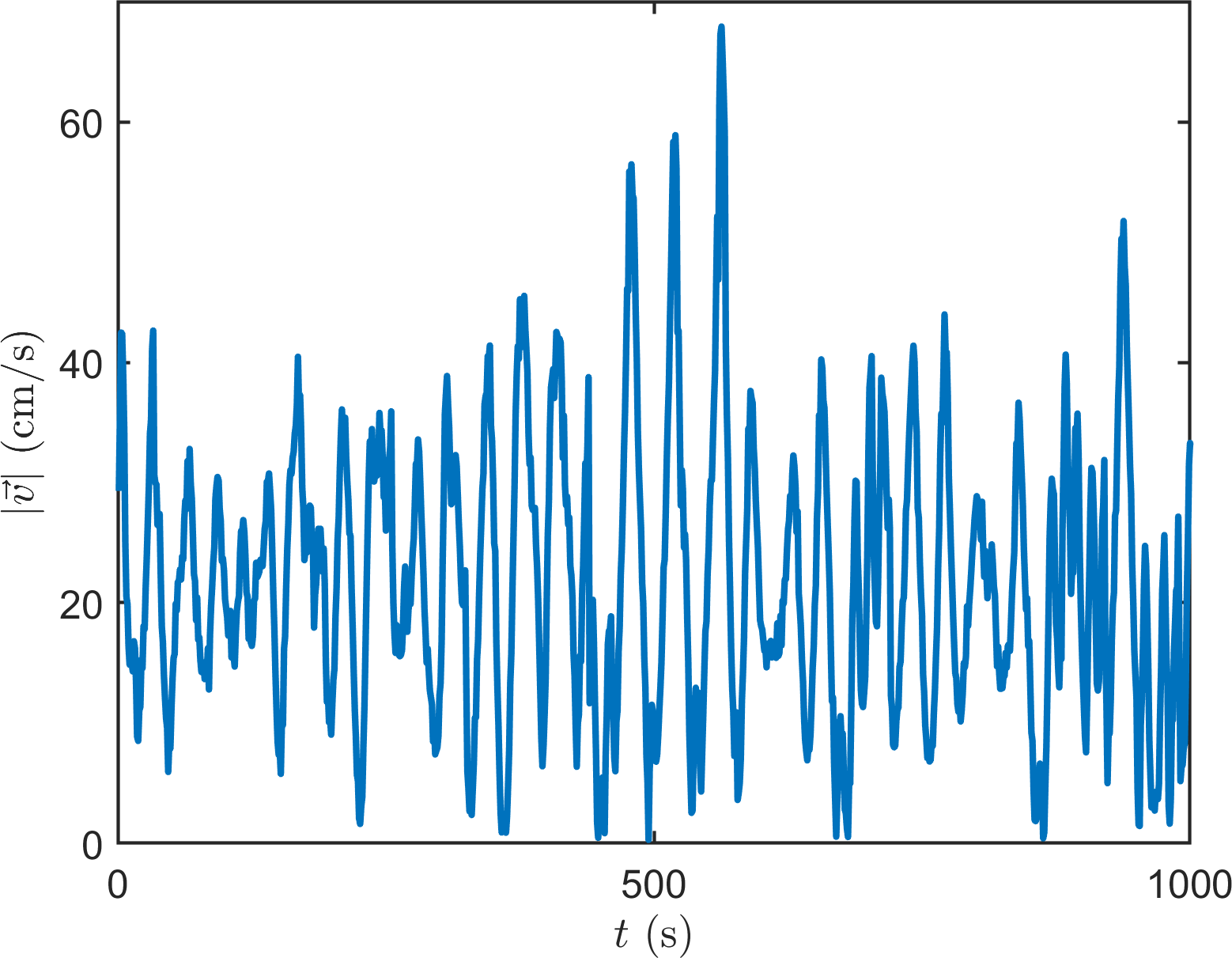}
\caption{\label{fig:linear_velocity} Linear velocity of the ball during a typical experimental run.}
\end{figure}

\subsection{Measurement of linear velocity}
During a typical experiment, the ``Squigglebot" is placed on a horizontal surface (optical table) and its motor is turned on wirelessly. To detect its position, a usb camera (Pixelink-PL-D734CU-T) \cite{pixelink} is fixed on top of the arena and the images are taken at 25 fps. In the case of a single ball, by tracking its position, one can compute its linear velocity. Whereas for experiments with multiple balls, one can construct the trajectories of each ball by using some kind of tracking algorithms \cite{crocker1996methods} and then calculate their linear velocities. Figure~\ref{fig:img} shows an image of a single moving ball during an experimental run. The ball is painted black for the ease of detection. The linear velocity of a single ball as a function of time is shown in Fig.~\ref{fig:linear_velocity}.


\begin{figure*}
\includegraphics[width=.9\linewidth]{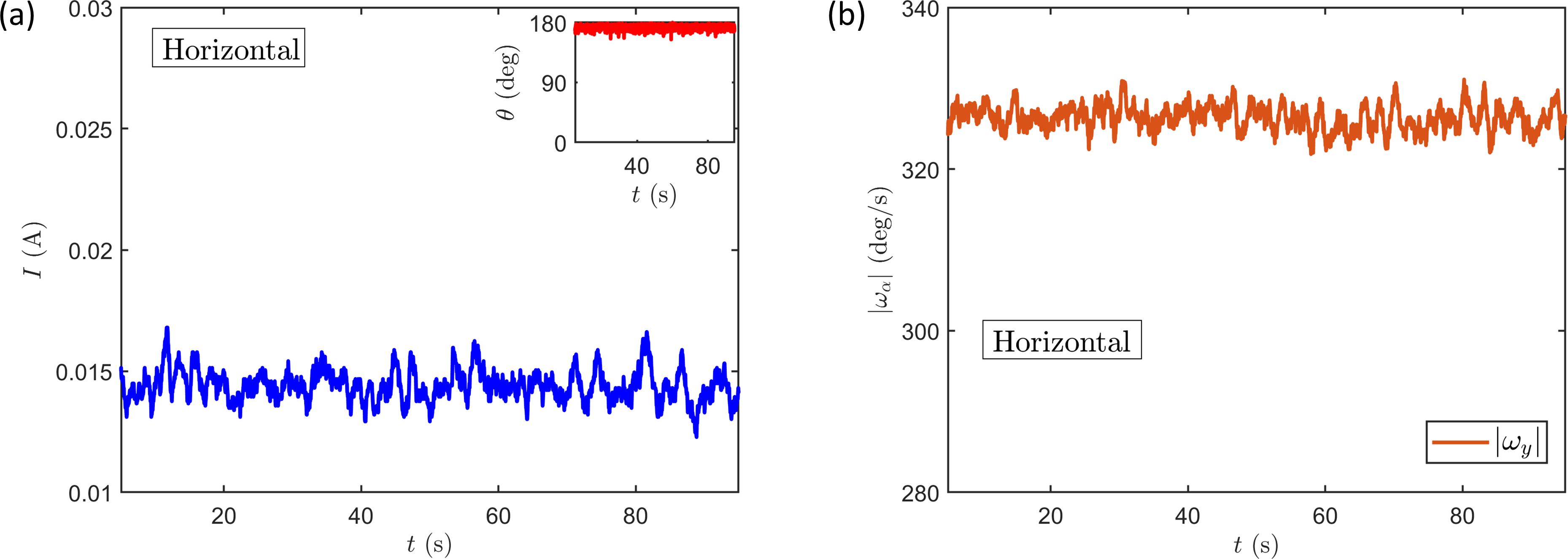}
\caption{\label{fig:horizontal} (a)  Time series of the current drawn by the motor in horizontal orientation. The angle $\theta$ is close to $180^{\circ}$ as shown in the inset. (b) Time series of $|\omega_y|$. $\omega_x$ and $\omega_z$ remain close to zero, hence not shown.}
\end{figure*}

\begin{figure*}
\includegraphics[width=.9\linewidth]{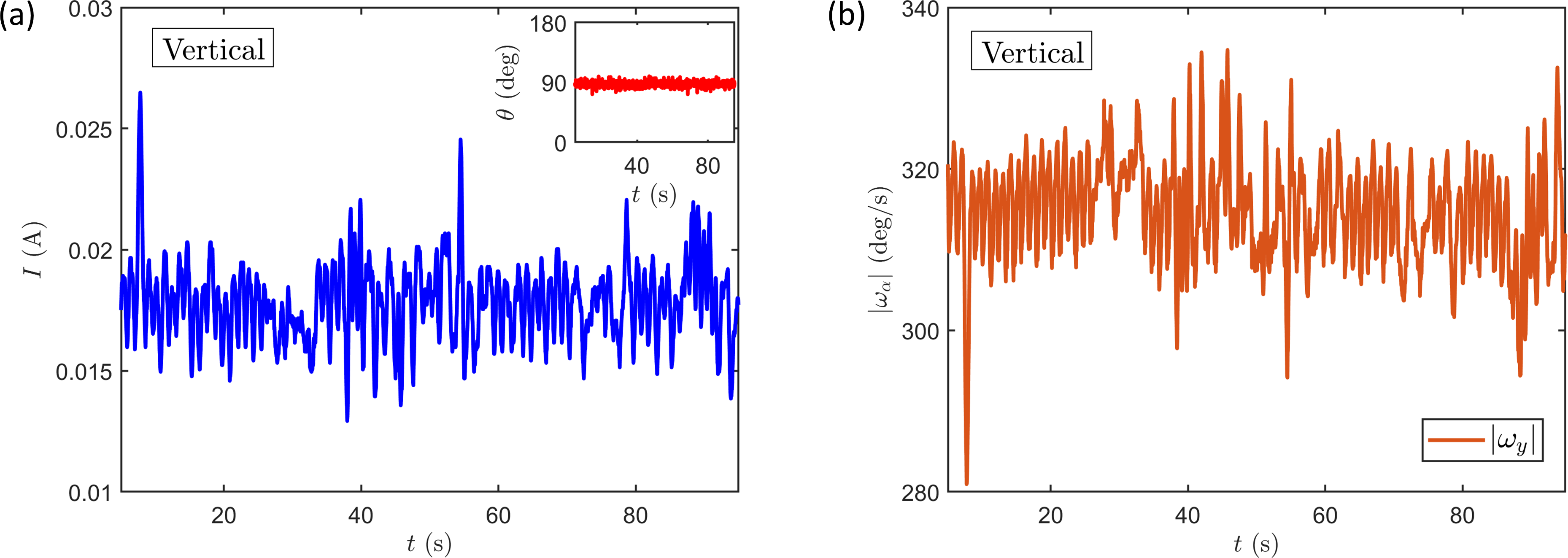}
\caption{\label{fig:vertical} (a) Time series of the current drawn by the motor in vertical orientation. The angle $\theta$ is close to $90^{\circ}$ as shown in the inset. (b) Time series of $|\omega_y|$. $\omega_x$ and $\omega_z$ remain close to zero, hence not shown.}
\end{figure*}

\begin{figure*}
\includegraphics[width=.9\linewidth]{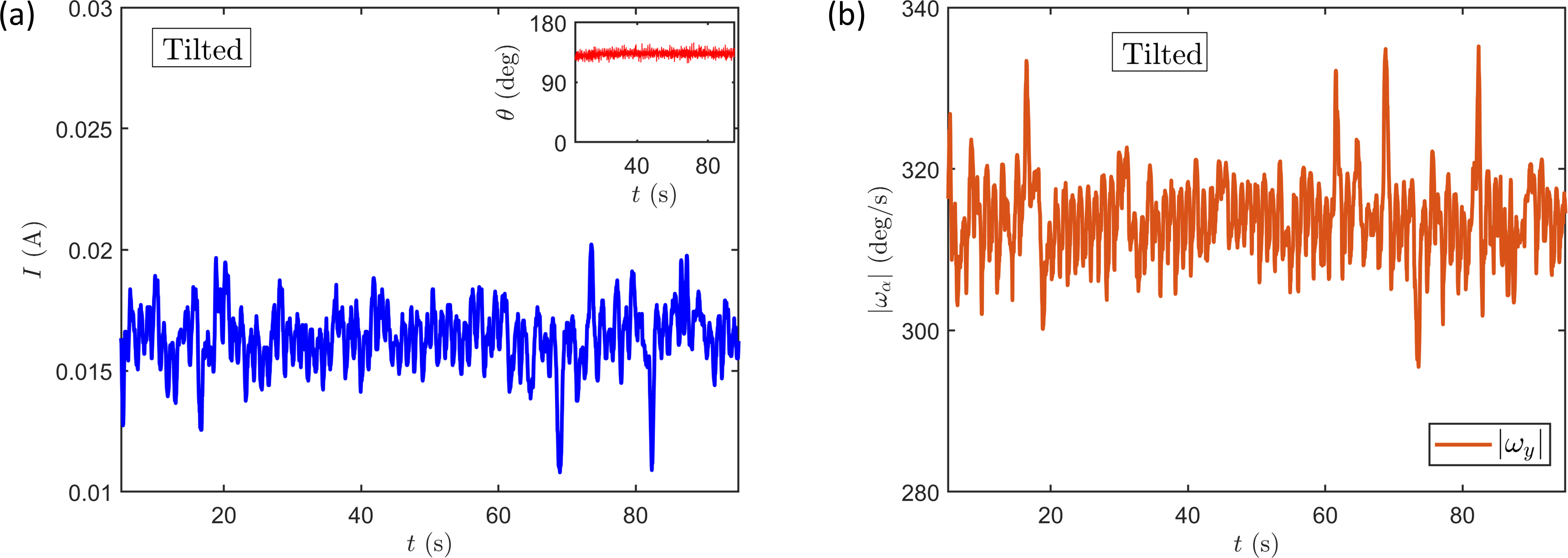}
\caption{\label{fig:tilted} (a) Time series of the current drawn by the motor in tilted orientation. The angle $\theta$ is shown in the inset which is close to $130^{\circ}$. (b) Time series of $|\omega_y|$. $\omega_x$ and $\omega_z$ remain close to zero, hence not shown.}
\end{figure*}

\begin{figure*}
\includegraphics[width=.9\linewidth]{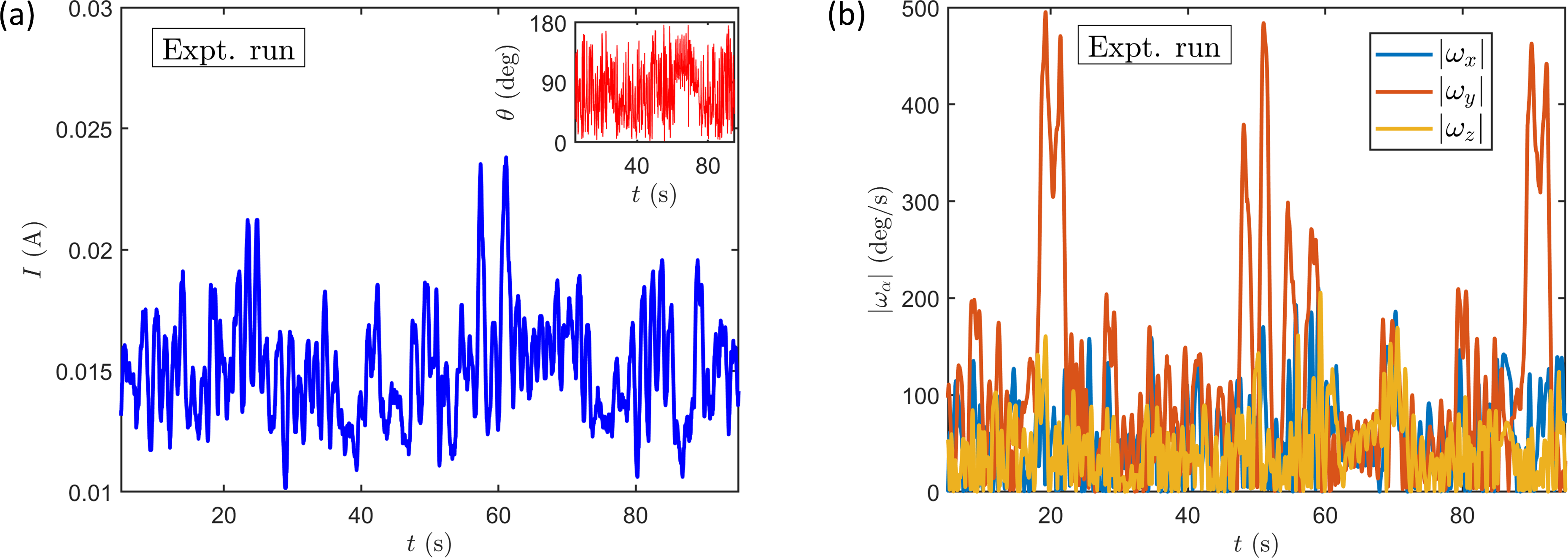}
\caption{\label{fig:bulk_run2} (a) Time series of the current drawn by the motor during a typical experimental run. The orientation of the plane of rotation is shown in the inset. (b) Time series of the magnitude of the different components of the angular velocity.}
\end{figure*}

\subsection{Measurement of the current drawn by the motor and its angular velocity}
As stated in \ref{sec:Design}~A, the brass load rotates along with the motor. If the plane of rotation is not horizontal, the motor has to work against the gravity to pull the load up and when the load comes down, motor is aided by the gravity. Hence, the current drawn by the motor changes in time depending on the angular position of the load in the plane of rotation as well as the orientation of the plane of rotation itself. In the present setup, the motor rotates about the body Y-axis which is perpendicular to the plane of rotation (Fig.~\ref{fig:mech_assembly}). Thus, the orientation is measured in terms of the angle between the gravity vector and body Y-axis as $\theta = \cos^{-1}\left(a_y/\sqrt{a_x^2+a_y^2+a_z^2}\right)$. Before presenting the data during a typical experimental run where the ball moves on a horizontal surface, we want to study the effects of the orientation of the plane of rotation on the current drawn by the motor as well as its angular velocities. This is achieved by holding the outer shell of the ball fixed in different orientations, such that the body of the motor as well as the load rotates along with it. The results are described in the next three sections.





\subsubsection{Horizontal rotation}
The motor rotates in the horizontal plane. Hence, the gravity vector and the axis of rotation is parallel to each other, i.e., $\theta \approx 180^{\circ}$ as can be seen from the inset of Fig.~\ref{fig:horizontal} (a). The corresponding time series of the current drawn by the motor is shown in Fig.~\ref{fig:horizontal} (a). The magnitude of the y-component of the angular velocity $|\omega_y|$ is shown in Fig.~\ref{fig:horizontal} (b). $\omega_x$ and $\omega_z$ components are not shown as they are close to zero.

\subsubsection{Vertical rotation}
Here the motor rotates in the vertical plane. Hence, the gravity vector and the axis of rotation is perpendicular to each other, i.e., $\theta \approx 90^{\circ}$ (Fig.~\ref{fig:vertical} (a) inset). The corresponding time series of the current drawn by the motor is shown in Fig.~\ref{fig:vertical} (a). Clearly the average current as well as the variation in current is higher as compared to the case of horizontal rotation due to the effect of gravity. The gravitational force $mg$ produces a torque which opposes and aids the motor torque during upward and downward motion respectively. For the same reason, $|\omega_y|$ shows a large variation in time which can be seen in Fig.~\ref{fig:vertical} (b). As before, $\omega_x$ and $\omega_z$ components are not shown as they are close to zero.

\subsubsection{Tilted rotation}
Here the motor rotates in a plane that is tilted at an angle with respect to the horizontal ($\theta \approx 130^{\circ}$). This is shown in Fig.~\ref{fig:tilted} (a) inset. The corresponding time series of the current drawn by the motor is shown in Fig.~\ref{fig:tilted} (a). Here, the average current as well as the variation in current is higher than that of the horizontal rotation but less than that of the vertical rotation. This is due to the fact that in the plane of rotation the gravitational force has a component $mg \cos\theta$ acting on the load which is less than $mg$. Figure~\ref{fig:tilted} (b) shows the time series of $|\omega_y|$. As before, $\omega_x$ and $\omega_z$ components are not shown as they are close to zero.

\subsubsection{Experimental run}
During the experiment, the ``Squigglebot" moves on a horizontal surface. Since there are no external torque acting on the ball, the outer shell and the brass load - motor body assembly both rotate in opposite direction to each other. 
Since the coupling between the outer shell of the ball and the motor shaft is not perfectly rigid, the ball has wobbling motion. The load can also fall in a direction perpendicular to the plane of rotation causing the ball to topple. Thus the plane of rotation has access to different angles between $0^{\circ}~\rm{to}~180^{\circ}$ as can be seen from Fig.~\ref{fig:bulk_run2} (a) inset. The corresponding time series of the current drawn by the motor is shown in Fig.~\ref{fig:bulk_run2} (a). Clearly all the three components of angular velocity are non-zero here which can be seen from Fig.~\ref{fig:bulk_run2} (b). 

\subsection{Software}
The source code for controlling the micro-controller, collecting data from the IMU and current sensors as well as transmitting data over Wi-Fi is written in the Arduino IDE. After booting up, ESP-12F connects to a local Wi-Fi network whose SSID and password is put in the code. It then starts a HTTP web server on port 80 and a Telnet server on port 23. The web server is used for the following jobs - (i) it is used to send the instruction for starting the measurement; (ii) it is used to toggle the state of the GPIO2 pin thereby turning the motor on or off; (iii) it is used to upload the source code to the micro-controller over Wi-Fi (OTA or over the air update) and (iv) it is used to reboot the micro-controller. When ESP-12F receives the instruction to start the measurement, it puts the GPIO2 pin to logic High and turns the motor on. It then starts collecting data from the IMU sensor over $\rm{I}^2 \rm{C}$ and also from the current sensor. The data is transferred to a computer on the same local Wi-Fi network using Telnet protocol which runs over the Transmission Control Protocol (TCP) at $25~\rm{Hz}$.

The source codes for collecting the sensor data over Wi-Fi from the micro-controller as well as the images from camera over USB are both written in Python. In order to synchronize the sensor and image data, multirun plugin of the Pycharm IDE is used to start both the codes simultaneously \cite{multirun}. It has been observed that during a typical experimental run, both the sensor and image data stay almost synchronized to each other - only about 20 secs off after 2 h 15 min. If one wants to get the simultaneous data from multiple ``Squigglebots", then they can be executed parallelly with threads using ThreadPoolExecutor subclass of the concurrent.futures module \cite{concurrent.futures}. All the source codes are provided in Github \cite{squigglebot_2022}.

\section{\label{sec:Conclusion}Conclusion}
In this paper, we have presented the design of an artificial active matter particle based on a battery operated rolling sphere with asymmetric mass distribution. Along with the output energy, the input energy can also be measured directly which is a non-trivial feature of this system since direct measurement of input energies at the level of individual active matter particles has not been done in the experiments previously. The mechanical construction as well as the details of electronic instrumentation based on open source hardware are provided, so are the source codes required to collect the data. We believe this can serve as a prototype system to study a variety of problems in non-equilibrium statistical physics, specially where details of the energetics are required experimentally. 

\bibliographystyle{apsrev4-1}
\bibliography{aipsamp}

\end{document}